\documentstyle[12pt]{article}
\begin{document}
\begin{flushright}
Preprint DFPD 96/TH/41\\ 
hep-th/9607171\\
July, 1996
\end{flushright}

\begin{center}
{\large\bf Comment on Covariant Duality Symmetric Actions}

\vspace{1cm}
{\bf Paolo Pasti}\footnote{e--mail: pasti@pd.infn.it},
{\bf Dmitrij Sorokin\footnote{on leave from Kharkov Institute of
Physics and Technology, Kharkov, 310108, Ukraine.\\~~~~~e--mail:
sorokin@pd.infn.it} and  
Mario Tonin\footnote{e--mail: tonin@pd.infn.it}
}

\vspace{0.5cm}
{\it Universit\`a Degli Studi Di Padova,
Dipartimento Di Fisica ``Galileo Galilei''\\
ed INFN, Sezione Di Padova,
Via F. Marzolo, 8, 35131 Padova, Italia}

\vspace{1.cm}
{\bf Abstract}
\end{center}
We demonstrate that an action proposed by A. Khoudeir and N. R. Pantoja 
in {\sl Phys. Rev.} {\bf D53}, 5974 (1996)
for endowing Maxwell theory 
with manifest electric--magnetic duality symmetry contains, besides the 
Maxwell field, additional propagating vector degrees of freedom. Hence 
it cannot be considered as a duality symmetric action for a {\it single} 
abelian gauge field.


\bigskip
The action 
proposed in \cite{kp} to describe abelian vector fields 
in four--dimensional Minkowski space
has the following form\footnote{for details of notation and convention 
see \cite{kp}}:
\begin{equation}
I= -\frac{1}{2}\int d^4x \ (u_n \ {\cal F}^{\alpha mn}
{\Phi}^{\alpha}_{mp}u^p + {\Lambda}^{\alpha mp}
{\Phi}^{\alpha}_{mp}),
\label{action}
\end{equation}
where $\alpha=1,2$, ${\cal L}_{\alpha\beta}$ is the antisymmetric unit 
tensor,
\begin{equation}\label{phi}
\Phi^{\alpha}_{mp} \equiv F^{\alpha}_{mp}
+{\cal L}^{\alpha\beta}{\cal F}^{\beta}_{mp},
\end{equation}
is a self--dual tensor $\Phi^{\alpha}_{mn} 
\equiv \frac{1}{2} \varepsilon_{mnpq}{\cal L}_{\alpha
\beta}{\Phi}^{\beta pq}$ constructed out of the field strengths of 
two abelian gauge fields $A^{\alpha}_m$ 
\begin{equation}
F^{\alpha}_{mn}= \partial_mA^{\alpha}_n-\partial_nA^{\alpha}_m,
\quad
{\cal F}^{\alpha mn}= \frac{1}{2} \varepsilon^{mnpq}
F^{\alpha}_{pq};
\label{dual}
\end{equation}
$u_m(x)$ is an auxiliary vector field satisfying the condition
\begin{equation}
u_{m} u^m=-1,
\label{norm}
\end{equation}
and
\begin{equation}\label{l}
{\Lambda}^{\alpha}_{mn} \equiv -\frac{1}{2} \varepsilon_{mnpq}
{\cal L}_{\alpha \beta}{\Lambda}^{\beta pq}
\end{equation}
 is anti--self--dual Lagrange multiplier, since $\Phi^{\alpha}_{mn}$
(\ref{phi}) is self--dual.  

The equations of motion one gets from (\ref{action}) reduce to
\begin{equation}
\frac{\delta}{\delta \Lambda^{\alpha}_{mn}} I = 0
\quad \Rightarrow \quad {\Phi}^{\alpha}_{mn} = 0,
\label{link}
\end{equation}
\begin{equation}
\frac{\delta}{\delta A^{\alpha}_m} I = 0 \quad \Rightarrow \quad
\varepsilon^{mnpq}\partial_p
{\Lambda}^{\alpha}_{nq} = 0.
\label{E-L}
\end{equation}
From (\ref{link}) it follows \cite{z,ss} that the field strength of one 
of the gauge fields $A^{\alpha}_m$ is dual to another one. Thus on the mass 
shell only one of $A^{\alpha}_m$ remains independent and the latter 
satisfies the free Maxwell equations of motion (see \cite{z,ss} for 
details).

At the same time
the general solution of Eq. (\ref{E-L}) is
\begin{equation}\label{b}
{\Lambda}^{\alpha}_{mn}=\partial_{[m}B^\alpha_{n]},
\end{equation}
where $B^\alpha_{n}(x)$ are vector fields which, because of 
anti--self--duality of ${\Lambda}^{\alpha}_{mn}$ (\ref{l}), satisfy
the Maxwell equations
\begin{equation}\label{m}
\partial^m\partial_{[m}B^\alpha_{n]}=0.
\end{equation}
Eq. (\ref{m}) is the point which demonstrates that the statement of Ref. 
\cite{kp} that on the mass shell ${\Lambda}^{\alpha}_{mn}=0$ fails. 
It might be so if the action (\ref{action}) had a local symmetry under 
which ${\Lambda}^{\alpha}_{mn}$ transformed as
\begin{equation}\label{sy}
\delta{\Lambda}^{\alpha}_{mn}=\partial_{[m}\phi^\alpha_{n]}-{\cal 
L}^{\alpha\beta}\varepsilon_{mn}^{~~~pq}\partial_{[p}\phi^\beta_{q]}
\end{equation}
with a vector parameter $\phi^\alpha_{n}(x)$. Then on the mass shall 
one might use this symmetry to eliminate $B^\alpha_{n}$.
[Note that simpler transformations of 
${\Lambda}^{\alpha}_{mn}$ of the form 
$\delta{\Lambda}^{\alpha}_{mn}=\partial_{[m}\phi^\alpha_{n]}$,
cannot be considered as a nontrivial 
local symmetry of the model since they leave the action invariant
only if  $\phi^\alpha_{n}(x)$ {\it a priori} (because of anti--self--duality 
of ${\Lambda}^{\alpha}_{mn}$) 
satisfies the dynamical Maxwell equations the same as 
${\Lambda}^{\alpha}_{mn}$ on the mass shell (\ref{m}). The action of any 
theory possesses such kind of trivial invariance].
One can see that if other fields of the model are inert under 
transformations with $\phi^\alpha_{n}(x)$
the action is not invariant 
under (\ref{sy}). 
Thus one should try to find appropriate 
transformations of $u_m$ and $A_m^\alpha$ which would cancel that of
${\Lambda}^{\alpha}_{mn}$ in the action. An argument against the 
existence of such transformations is that for the local symmetry to 
be present there must be first--class constraints on dynamical variables of 
the model which generate this symmetry. However there are no relevant 
constraints in the case at hand. Analogous situation takes place in 
simpler case of chiral bosons \cite{si,fj,sr} where an action proposed 
in \cite{sr} has a Lagrange multiplier term  linear in derivatives of 
physical fields (like in (\ref{action})). There the Lagrange multiplier is a 
propagating degree of freedom which causes the problem with 
unitarity of the model of \cite{sr} (see \cite{b} for detailed 
discussion of these points).

By the same reasons the model of \cite{kp} contains additional 
propagating vector degrees of freedom $B^\alpha_{n}$ and cannot be 
considered as a covariant version of a duality--symmetric free Maxwell 
action \cite{z,ss}.
A consistent Lorentz covariant way of constructing duality--symmetric actions
was proposed in \cite{du}, and alternative formulations, based on an 
infinite number of auxiliary fields, were considered recently in \cite{inf}.

\end{document}